\documentclass[conference]{IEEEtran}
\ifCLASSINFOpdf
\else
\fi
\usepackage[cmex10]{amsmath}
\ifCLASSOPTIONcompsoc
  \usepackage[caption=false,font=normalsize,labelfont=sf,textfont=sf]{subfig}
\else
  \usepackage[caption=false,font=footnotesize]{subfig}
\fi
\usepackage{fixltx2e}
\newcommand{\matr}[1]{\mathbf{#1}}
\DeclareMathOperator{\diag}{diag}
\DeclareMathOperator{\blkdiag}{blkdiag}
\DeclareMathOperator{\hermit}{H}
\renewcommand{\Re}{\operatorname{Re}}
\renewcommand{\Im}{\operatorname{Im}}

\usepackage{bm}
\usepackage{stfloats}

\usepackage{pgfplots}
\usetikzlibrary{shapes}
\pgfplotsset{compat=newest}
\newlength\figureheight	
\newlength\figurewidth

\newcommand{\columnplot}{
\setlength\figureheight{0.24\textwidth}
\setlength\figurewidth{0.37\textwidth}
}	

\newcommand{\columnplottwo}{
\setlength\figureheight{0.22\textwidth}
\setlength\figurewidth{0.37\textwidth}
}

\usepackage[exponent-product = \cdot]{siunitx}

\begin{document}
%
\title{Channel Parameter Estimation for LOS MIMO Systems}

\author{\IEEEauthorblockN{Tim H{\"a}lsig and Berthold Lankl}
\IEEEauthorblockA{Institute for Communications Engineering\\
Universit{\"a}t der Bundeswehr M{\"u}nchen, Germany\\
Email: tim.haelsig@unibw.de}
}


%


\maketitle

\begin{abstract}
In this paper we consider the estimation of channel coefficients and frequency offsets for LOS MIMO systems. We propose that by exploiting the structure of the channel matrix, which is present due to the geometrical nature of the channel, the estimation process can be enhanced. If a single oscillator setup is used at transmitter and receiver, respectively, this structure is preserved and can be exploited. Some methods using this fact are discussed and their performance is evaluated with respect to estimation accuracy, revealing that with relatively short training sequences, estimation results close to the fundamental bounds can be achieved.
\end{abstract}


\let\thefootnote\relax\footnotetext{This work was supported in part by the German Research Foundation (DFG) in the framework of priority program SPP~1655 "Wireless Ultra High Data Rate Communication for Mobile Internet Access".}

%
\IEEEpeerreviewmaketitle

\section{Introduction}

Channel estimation and synchronization are fundamental tasks that need to be solved for every communications system in order to equalize the channel and provide the means for data transmission. For MIMO systems this can become very complex due to the possibly high number of parameters that need to be estimated depending on the system setup. Estimation of the MIMO channel coefficients and of carrier frequency offsets is well established for classical fading channels. The activity and progress in the field of millimeter wave circuits have made line-of-sight~(LOS) MIMO a prime technique for indoor and outdoor communications scenarios requiring high data rates. LOS MIMO is based on spherical wave modeling \cite{Calabro2003}, which generates a channel matrix that is highly dependent on the geometry of the antenna arrangements.

For fading channels there have been several investigations over the years determining fundamental limits and viable training schemes that allow the estimation of the channel, as well as frequency offsets. In \cite{Besson2003} narrowband Rayleigh fading channels in conjunction with different frequency offsets for each transmit antenna are investigated for pilot based schemes. The authors derive the Cram{\'{e}}r-Rao Bound~(CRB) for that case and propose a simplified maximum likelihood~(ML) and a correlation-based estimator coming close to that bound. An overview of training-based MIMO channel estimation can be found in \cite{Biguesh2006}. Several estimation methods are discussed regarding their performance and complexity and the subject of optimal training sequences for them is reviewed. The works in \cite{Ghogho2006} and \cite{Qu2006} investigate the training design for the estimation of MIMO channels, including frequency selective fading scenarios, with carrier frequency offsets. Suitable training sequences and different estimation methods with reasonable computational cost are suggested for different setups. Pilot-assisted frequency synchronization for massive MIMO systems is studied in \cite{Cheng2013}, where the channel gains are assumed to be known. The CRB and ML estimation are provided and an achievable rate analysis is carried out.

In this paper we will show some results on how to estimate channel coefficients and frequency offsets specifically for LOS channel MIMO systems, which has rarely been considered in the literature. In principle, the same techniques used in the literature can also be applied to LOS MIMO systems for parameter estimation. However, the inherent structure of the channel can be exploited in order to reduce estimation complexity or improve estimation accuracy.

Consider $(\cdot)^T$ and $(\cdot)^H$ to denote transpose and conjugate transpose, respectively. Boldface small letters, e.g., $\matr{x}$, are used for vectors while boldface capital letters, e.g., $\matr{X}$, are used for matrices. Furthermore, $\matr{I}_N$ is the $N\times N$ identity matrix, while $\diag\left(\matr{x}\right)$ and $\blkdiag\left(\matr{X}\right)$ denote the diagonal and block diagonal matrix with diagonal elements of vector $\matr{x}$ and matrix $\matr{X}$, respectively.

\newcounter{mytempeqncnt}
\begin{figure*}[!b]
\normalsize
\setcounter{mytempeqncnt}{\value{equation}}
\setcounter{equation}{9}

\hrulefill
\vspace*{0pt}

\begin{equation}
\label{eq:crb_fo}
\text{CRB}\left(\bm{\theta}_m\right)= \frac{\sigma^2}{2}
\begin{bmatrix}
\Re\{\matr{X}_{m,\omega}^{\hermit}\matr{X}_{m,\omega}\}& -\Im\{\matr{X}_{m,\omega}^{\hermit}\matr{X}_{m,\omega}\} & \Re\{\matr{X}_{m,\omega}^{\hermit}\matr{D}_m\}\\
\Im\{\matr{X}_{m,\omega}^{\hermit}\matr{X}_{m,\omega}\}& \Re\{\matr{X}_{m,\omega}^{\hermit}\matr{X}_{m,\omega}\} & \Im\{\matr{X}_{m,\omega}^{\hermit}\matr{D}_m\}\\
\Re\{\matr{D}_m^{\hermit}\matr{X}_{m,\omega}\}& -\Im\{\matr{D}_m^{\hermit}\matr{X}_{m,\omega}\} & \Re\{\matr{D}_m^{\hermit}\matr{D}_m\}\\
\end{bmatrix}^{-1}
\end{equation}

\setcounter{equation}{\value{mytempeqncnt}}
\vspace*{0pt}
\end{figure*}

\section{System Model}

Consider the narrowband received signal of a MIMO system in baseband to be defined by
\begin{equation}
y_m(t) = \sum_{n=1}^N h_{mn}\cdot x_n(t)\cdot e^{j2\pi\Delta f_{mn}t} e^{j\Delta \phi_{mn}} + n_m(t) \label{eq:multModel}
\end{equation}
where $n=1\ldots N$ and $m=1\ldots M$ describe the index and number of transmit and receive antennas. Furthermore, $h_{mn}$ is the channel coefficient between the corresponding antennas and $x_n(t)$ is the continuous information carrying waveform transmitted from the $n$th antenna in complex baseband representation. The frequency and phase differences between the different oscillators at transmitter and receiver are denoted as $\Delta f_{mn}$ and $\Delta \phi_{mn}$. The term $n_m(t)$ is additive noise with complex Gaussian distribution at the $m$th antenna. Note that the frequency offsets correspond to the normalized frequency value, i.e., $\Delta f_{mn}=\frac{f_n-f_m}{f_s}$ where $f_s$ is the symbol rate and $f_n$, $f_m$ are the frequencies of the corresponding oscillators at transmitter and receiver. 

For the case of a single oscillator at transmitter and receiver, respectively, this reduces to 
\begin{equation}
y_m(t) = e^{j2\pi\Delta ft} e^{j\Delta \phi}\cdot\sum_{n=1}^N h_{mn}\cdot x_n(t) + n_m(t) \label{eq:commonModel}
\end{equation}
which is generally easier to estimate and compensate since less parameters have to be considered. In practice, this setup might not always be realizable, e.g., due to a large number of antennas.

For a pure LOS channel \cite{Bøhagen2007,Halsig2015} the coefficients are dependent on the geometric setup of the antenna arrays, determined through
\begin{align}
h_{mn} &= a_{mn}\cdot \exp\left(-j2\pi f_n\cdot\tau_{mn}\right) \\
	&= a_{mn}\cdot \exp\left(-j2\pi \frac{r_{mn}}{\lambda_n}\right) \label{eq:hLOS}
\end{align}
where $a_{mn}$ is the corresponding attenuation coefficient and $\tau_{mn}$ is the propagation delay between antenna $n$ and antenna $m$, which is given by the distance between the antennas $r_{mn}$ and the wavelength of the $n$th transmit oscillator $\lambda_n=c/f_n$ where $c$ is the speed of light. The value of $a_{mn}$ should in a LOS scenario be approximately equal across the different paths and can thus be assumed constant for all $h_{mn}$ and be neglected for the further analysis. 

\section{Cram{\'{e}}r-Rao Bound}

The CRB offers the fundamental limit that an estimator can possibly achieve. To derive it first assume that $x_n(t)$ is now a training signal that is going to be used to estimate the unknown parameters of the channel. Using $P$ discrete samples of rate $f_s$, we can write the signal received at the $m$th antenna as a vector with
\begin{equation}
\matr{y}_m = \underbrace{\left(\matr{\Omega}_m\odot\matr{X}\right)}_{\matr{X}_{m,\omega}}\underbrace{\matr{\Phi}_m\matr{h}_m}_{\matr{h}_{m,\phi}} + \matr{n}_m \label{eq:y}
\end{equation}
where $\matr{y}_m=\left[y_m(1),\ldots,y_m(P)\right]^T$, $\matr{h}_{m}=\left[h_{m1},\ldots,h_{mN}\right]^T$ and $\matr{n}_m=\left[n_m(1),\ldots,n_m(P)\right]^T\sim\mathcal{C}\mathcal{N}(0,\sigma^2\matr{I}_{P})$. Further, the phase shift is in $\matr{\Phi}_m=\diag\left(e^{j\Delta\phi_{m1}},\ldots,e^{j\Delta\phi_{mN}}\right)$, $\odot$ is the Hadamard product, and
\begin{equation}
\matr{X}=
\begin{bmatrix}
x_1(1) & \cdots & x_N(1) \\
\vdots & \ddots & \vdots \\
x_1(P) & \cdots & x_N(P) \\
\end{bmatrix} 
=
\begin{bmatrix}
\matr{x}(1)^T \\
\vdots \\
\matr{x}(P)^T \\
\end{bmatrix}\text{,} \notag
\end{equation}
\begin{equation}
\matr{\Omega}_m=
\begin{bmatrix}
e^{j\Delta\omega_{m1}} & \cdots & e^{j\Delta\omega_{mN}} \\
e^{j2\Delta\omega_{m1}} & \cdots & e^{j2\Delta\omega_{mN}} \\
\vdots & \ddots & \vdots   \\
e^{jP\Delta\omega_{m1}} & \cdots & e^{jP\Delta\omega_{mN}} \\
\end{bmatrix} \text{,} \notag
\end{equation}
with $\Delta\omega_{mn}=2\pi\Delta f_{mn}$. Note that for the single oscillator setup, there is no dependence on $m$ and $n$ in the matrices $\matr{\Phi}_m$ and $\matr{\Omega}_m$.

We can also write the complete received vector as
\begin{equation}
\matr{y} = \matr{X}_\omega\matr{h}_\phi + \matr{n} \label{eq:y_comp}
\end{equation}
where  $\matr{y}=\left[\matr{y}_1^T,\ldots,\matr{y}_M^T\right]^T$, $\matr{h}_\phi=\left[\matr{h}_1^T\matr{\Phi}_1,\ldots,\matr{h}_M^T\matr{\Phi}_M\right]^T$, $\matr{n}=\left[\matr{n}_1^T,\ldots,\matr{n}_M^T\right]^T$, and the frequency offset impaired training matrix is in $\matr{X}_\omega=\blkdiag\left(\matr{\Omega}_1\odot\matr{X},\ldots,\matr{\Omega}_M\odot\matr{X}\right)$. As noted in other works \cite{Besson2003,Ghogho2006}, the estimation of the parameters for each of the receiving antennas is decoupled (fisher information matrix is block diagonal, CRB is block diagonal) and can be carried out independently, and hence we will focus in the following on \eqref{eq:y} rather than \eqref{eq:y_comp}. Note that we have merged the channel coefficients and the phase shifts into one term $\matr{h}_{m,\phi}$, this will be explained in the next section.

\subsection{No Frequency Offset}
Let us first investigate the case when $\Delta\omega_{mn}=0$. Then, the model reduces to
\begin{equation}
\matr{y}_m = \matr{X}\matr{h}_{m,\phi} + \matr{n}_m
\end{equation}
where the parameter vector to be estimated per receive antenna is $\bm{\theta}_m=\begin{bmatrix}\Re\{\matr{h}_{m,\phi}^T\} & \Im\{\matr{h}_{m,\phi}^T\}\end{bmatrix}^T$, and since $\matr{n}_m$ is a white Gaussian noise vector, the CRB is readily found \cite{Besson2003,Stoica2001} by
\begin{equation}
\text{CRB}\left(\bm{\theta}_m\right)= \frac{\sigma^2}{2}
\begin{bmatrix}
\Re\{\matr{X}^{\hermit}\matr{X}\}& -\Im\{\matr{X}^{\hermit}\matr{X}\} \\
\Im\{\matr{X}^{\hermit}\matr{X}\}& \Re\{\matr{X}^{\hermit}\matr{X}\}
\end{bmatrix}^{-1}  \label{eq:crb_nfo} \text{.}
\end{equation}

\subsection{Frequency Offset Impaired}
Using the full form also including the frequency offsets 
\begin{equation}
\matr{y}_m = \matr{X}_{m,\omega}\matr{h}_{m,\phi} + \matr{n}_m
\end{equation}
the new parameter vector of interest to be determined is $\bm{\theta}_m=\begin{bmatrix}\Re\{\matr{h}_{m,\phi}^T\} & \Im\{\matr{h}_{m,\phi}^T\} & \bm{\omega}_{m}^T\end{bmatrix}^T$, where the vector containing the frequency offsets is $\bm{\omega}_{m}=\left[\Delta\omega_{m1},\ldots,\Delta\omega_{mN}\right]^T$. The CRB is found, e.g., in \cite{Stoica2001} and repeated at the bottom of the page \eqref{eq:crb_fo}, with $\matr{D}_m = \diag\left(1,\ldots,P\right)\cdot\matr{X}_{m,\omega}\cdot\diag\left(\matr{h}_{m,\phi}\right)$.

\addtocounter{equation}{1}

\section{LOS MIMO}
As can be seen in \eqref{eq:hLOS}, the channel coefficients in the LOS MIMO case are determined by the distances between transmit and receive antennas $r_{mn}$. We write the channel coefficients including the phase shifts
\begin{align}
h_{mn,\phi} =\exp\left(-j2\pi \frac{r_{mn}}{\lambda_n}\right) \cdot \exp\left(j\Delta \phi_{mn}\right) \text{} \label{eq:chanphi}
\end{align}
which can be considered as one joint term, c.f. \eqref{eq:y}. This is possible because the introduced phase shifts, which do not vary in time, correspond to row and column operations on the initial $\matr{H}=\left[\matr{h}_1,\ldots,\matr{h}_M\right]^T$ which do not change the condition number of the matrix. The phase shifts will thus be fully compensated as part of a generic MIMO receiver by using the joint channel matrix $\matr{H}_{\phi}=\left[\matr{\Phi}_1\matr{h}_1,\ldots,\matr{\Phi}_M\matr{h}_M\right]^T$, whose entries are determined through \eqref{eq:chanphi}. Nevertheless, the phase shifts will have an impact on the estimation of the channel matrix as will be seen later.

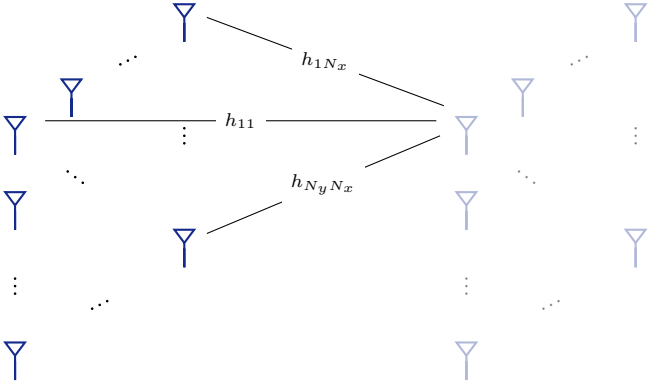
\begin{figure}[!t]
\centering
\def\antenna{%
   -- +(0mm,-2.5mm) -- +(0mm,0.75mm) -- +(1.31mm,2.5mm) -- +(-1.31mm,2.5mm) -- +(0mm,0.75mm)
}

\definecolor{mycolor1}{rgb}{0.0784313753247261,0.168627455830574,0.549019634723663}%
\definecolor{mycolor2}{rgb}{1,1,1}%

\tikzset{>=latex}

\begin{tikzpicture}[font=\scriptsize]

\draw[color=mycolor1,fill=mycolor2,thick] (0,0) \antenna;
\draw[color=mycolor1,fill=mycolor2,thick] (0,1) \antenna;

\draw[fill=black] (0,-1) circle (0.1mm);
\draw[fill=black] (0,-0.9) circle (0.1mm);
\draw[fill=black] (0,-1.1) circle (0.1mm);

\draw[color=mycolor1,fill=mycolor2,thick] (0,-2) \antenna;

\draw[color=mycolor1,fill=mycolor2,thick] (0.75,1.5) \antenna;

\draw[fill=black] (1.5,2) circle (0.1mm);
\draw[fill=black] (1.4,1.95) circle (0.1mm);
\draw[fill=black] (1.6,2.05) circle (0.1mm);

\draw[fill=black] (1.125,-1.25) circle (0.1mm);
\draw[fill=black] (1.025,-1.3) circle (0.1mm);
\draw[fill=black] (1.225,-1.2) circle (0.1mm);

\draw[color=mycolor1,fill=mycolor2,thick] (2.25,2.5) \antenna;

\draw[color=mycolor1,fill=mycolor2,thick] (2.25,-0.5) \antenna;

\draw[fill=black] (2.25,1) circle (0.1mm);
\draw[fill=black] (2.25,1.1) circle (0.1mm);
\draw[fill=black] (2.25,0.9) circle (0.1mm);

\draw[fill=black] (0.8,0.45) circle (0.1mm);
\draw[fill=black] (0.7,0.525) circle (0.1mm);
\draw[fill=black] (0.9,0.375) circle (0.1mm);

\begin{scope}[xshift=6cm, opacity=0.33]

\draw[color=mycolor1,fill=mycolor2,thick] (0,0) \antenna;
\draw[color=mycolor1,fill=mycolor2,thick] (0,1) \antenna;

\draw[fill=black] (0,-1) circle (0.1mm);
\draw[fill=black] (0,-0.9) circle (0.1mm);
\draw[fill=black] (0,-1.1) circle (0.1mm);

\draw[color=mycolor1,fill=mycolor2,thick] (0,-2) \antenna;

\draw[color=mycolor1,fill=mycolor2,thick] (0.75,1.5) \antenna;

\draw[fill=black] (1.5,2) circle (0.1mm);
\draw[fill=black] (1.4,1.95) circle (0.1mm);
\draw[fill=black] (1.6,2.05) circle (0.1mm);

\draw[fill=black] (1.125,-1.25) circle (0.1mm);
\draw[fill=black] (1.025,-1.3) circle (0.1mm);
\draw[fill=black] (1.225,-1.2) circle (0.1mm);

\draw[color=mycolor1,fill=mycolor2,thick] (2.25,2.5) \antenna;

\draw[color=mycolor1,fill=mycolor2,thick] (2.25,-0.5) \antenna;

\draw[fill=black] (2.25,1) circle (0.1mm);
\draw[fill=black] (2.25,1.1) circle (0.1mm);
\draw[fill=black] (2.25,0.9) circle (0.1mm);

\draw[fill=black] (0.8,0.45) circle (0.1mm);
\draw[fill=black] (0.7,0.525) circle (0.1mm);
\draw[fill=black] (0.9,0.375) circle (0.1mm);

\end{scope}

\draw[-] (0.4,1.2) -- (5.6,1.2) node[pos=0.5,fill=white] {$h_{11}$};
\draw[-] (2.55,2.575) -- (5.7,1.4) node[pos=0.5,fill=white] {$h_{1N_x}$};
\draw[-] (2.55,-0.3) -- (5.65,1.0) node[pos=0.5,fill=white] {$h_{N_y N_x}$};

\end{tikzpicture}
\caption{Example of a symmetric uniform rectangular array setup generating a channel matrix with block Toeplitz structure.}
\label{fig:setup}
\end{figure}

Due to the geometrical structure of the channel, the channel coefficients from one receive antenna to another typically do not vary randomly as is the case for Rayleigh scattering. For example, for all of the optimal symmetric ($M=N$) uniform array designs, e.g., \cite{Bøhagen2007,Halsig2015,Song2015}, the matrix $\matr{H}$ will have a block Toeplitz structure, i.e., 
\begin{equation}
\matr{H}=
\begin{bmatrix}
\matr{H}_1 & \matr{H}_2 & \cdots & \matr{H}_{N_y} \\
\matr{H}_2 & \matr{H}_1 & \cdots & \matr{H}_{N_y-1} \\
\matr{H}_3 & \matr{H}_2 & \cdots & \matr{H}_{N_y-2}   \\
\vdots & \vdots & \ddots & \vdots   \\
\matr{H}_{N_y} & \matr{H}_{N_y-1} & \cdots & \matr{H}_1 \\
\end{bmatrix} \text{,} \notag
\end{equation}
where $N_y$ is the number of elements in the first array dimension and the sub-matrices of index $n_y$ have Toeplitz structure with 
\begin{equation}
\matr{H}_{n_y}=
\begin{bmatrix}
{h}_{n_y1} & {h}_{n_y2} & \cdots & {h}_{n_yN_x} \\
{h}_{n_y2} & {h}_{n_y1} & \cdots & {h}_{n_yN_x-1} \\
{h}_{n_y3} & {h}_{n_y2} & \cdots & {h}_{n_yN_x-2}   \\
\vdots & \vdots & \ddots & \vdots   \\
{h}_{n_yN_x} & {h}_{n_yN_x-1} & \cdots & {h}_{n_y1} \\
\end{bmatrix} \text{,} \notag
\end{equation}
where $N_x$ is the number of elements in the second array dimension and $N=N_y\cdot N_x$. Note that there are $N_y$ different sub-matrices with $N_x$ different entries. An example of such a setup is given in Fig.~\ref{fig:setup}. For the case of uniform~linear~arrays~(ULAs), a special case where $N_x=1$ or $N_y=1$, the matrix reduces to standard Toeplitz form, i.e., $\matr{h}_1,\ldots,\matr{h}_{M}$ are circularly shifted versions of each other.

In general, for every regular shaped antenna arrangement, including non-symmetric cases, there will be a part of the matrix that has block Toeplitz structure and a part that depends more specifically on the chosen arrangement and alignment of the arrays. We will in this paper focus on symmetric uniform~rectangular~array~(URA) setups as they deliver the highest capacity with the lowest form factor for pure LOS channels \cite{Song2015}.

\subsection{Estimation in Frequency Offset free Case}
Any training matrix $\matr{X}$ having orthogonal columns under transmit power constraint is optimal in the sense that it minimizes the CRB \cite{Biguesh2006}, i.e., $\matr{X}^{\hermit}\matr{X}=\matr{I}_N$. Note that this requires a pilot sequence of length $P\geq N$ samples.

First, let us consider the URA arrangement mentioned above generating a block Toeplitz structure in a single oscillator setup, as in \eqref{eq:commonModel}, without frequency offset. In that case the joint channel matrix $\matr{H}_{\phi}$ will also be of block Toeplitz character and we can use that information to infer the full channel matrix from one transmitted training vector $\matr{x}(p)$ ($P=1$) and its received vector. In doing so we gain one row/column $\matr{h}_{m,\phi}$ of the matrix which is sufficient to build up the complete channel matrix. Using more training vectors can in this case be useful to improve the accuracy of the estimates by, for example, averaging the corresponding entries of two rows to reduce the impact of the Gaussian noise.

Now let us look at the case where there are multiple oscillators \eqref{eq:multModel}. In general, a longer training sequence is needed because the Toeplitz structure is completely obscured by introducing $M\cdot N$ unknown independent phase shifts and thus we may resort to the methods discussed in \cite{Biguesh2006}.

\subsection{Estimation in Frequency Offset corrupted Case}
The case including frequency offset estimation is more complex and has been discussed various times in the literature. It was shown that maximum likelihood estimators can be used to achieve the CRBs in a Rayleigh channel case for such a setup \cite{Besson2003}, but requiring a high computational complexity. Those estimators are based on the premise that $M\cdot N$ random channel coefficients and frequency offsets have to be estimated.

We start again with the case of a single oscillator as in \eqref{eq:commonModel}. Then, for the LOS MIMO case we can exploit the non-randomness of the channel coefficients similarly to the previous section. 
Assume that the $P$ training vectors are given by $\matr{x}(p)=\left[1,0,\ldots,0\right]^T$ and circularly shifted versions thereof, which results in the easiest case to the training matrix 
\begin{equation}
\matr{X}=
\begin{bmatrix}
1 & 0 & \cdots & 0\\
0 & 1 & \cdots & 0\\
\vdots & \vdots & \ddots & \vdots\\
0 & 0 & \cdots & 1\\
\end{bmatrix} \text{} \notag
\end{equation}
of dimensions $P\times N$.
We now focus on ULAs, namely standard Toeplitz structure $\matr{H}=\matr{H}_1$, for the sake of clarity. The received vectors after $P$ pilots for the first two receive antennas will contain the following
\begin{equation}
\matr{y}_1=
\begin{bmatrix}
e^{j\Delta\omega}h_{11,\phi} \\
e^{j2\Delta\omega}h_{12,\phi} \\
\vdots \\
e^{jP\Delta\omega}h_{1N,\phi}
\end{bmatrix} \text{, }
\matr{y}_2=
\begin{bmatrix}
e^{j\Delta\omega}h_{12,\phi} \\
e^{j2\Delta\omega}h_{11,\phi} \\
\vdots \\
e^{jP\Delta\omega}h_{1N-1,\phi}
\end{bmatrix} \text{.} \notag
\end{equation}
Note that the frequency offset part stays the same throughout the receive vectors $\matr{y}_m$ while the channel coefficients follow the Toeplitz structure. It should be visible that there are numerous entries of the received vectors from different antennas, which can be used to eliminate the impact of channel coefficients and gain an estimate of the frequency offset $\Delta\omega$. One example using $\matr{y}_1$ and $\matr{y}_2$ from above yields
\begin{align}
e^{j\Delta\hat{\omega}}=\frac{e^{j2\Delta\omega}h_{11,\phi}}{e^{j\Delta\omega}h_{11,\phi}}&=\frac{e^{j3\Delta\omega}h_{12,\phi}}{e^{j2\Delta\omega}h_{12,\phi}}=\ldots \notag \\ 
&=\frac{e^{jP\Delta\omega}h_{1N-1,\phi}}{e^{j(P-1)\Delta\omega}h_{1N-1,\phi}}=e^{j\Delta\omega} \label{eq:freq_est_example}
\end{align}
where $\Delta\hat{\omega}$ is the estimated frequency offset. Note that in this case two training vectors, $P=2$, are sufficient to gain multiple estimates of the frequency offset, which can in turn be used to obtain channel estimates. As before, additional training vectors can be used to improve the estimates.

For the multiple oscillator setup, the same problem of losing the Toeplitz structure occurs and more training vectors are needed. Furthermore, there are $M\cdot N$ independent frequency offsets which need to be estimated and additionally obscure the initial structure of the matrix, we may use \cite{Besson2003,Qu2006}.

\subsection{Comments on Estimators and Optimality}
From \eqref{eq:freq_est_example} we can write a simple estimator for the frequency offset of ULAs ($M=N$) that uses the received vectors from two neighboring antennas as
\begin{equation}
\Delta\hat{\omega} = \frac{1}{M/2\cdot (P-1)} \sum_{m=1}^{M/2}\sum_{p=1}^{P-1} \arg\left\{\frac{y_{2m}(p+1)}{y_{2m-1}(p)}\right\} \label{eq:freq_est} \text{}
\end{equation}
which works if the number of antennas is even. As an example we can also give an estimator for the diagonal entries of the channel matrix for this case as
\begin{equation}
\hat{h}_{mm,\phi} = \frac{1}{P} \sum_{m=p=1}^{P} e^{-jp\Delta\hat{\omega}}y_{m}(p) \text{.}
\end{equation}
Generally, we can choose among many different options when it comes to determining which of the entries and how they should be used for the estimation of frequency offsets, as well as channel coefficients. The performance of the different options will mostly depend on how the Toeplitz structure is exploited and how distinct it is throughout the matrix. Therefore, note that none of the estimators used in this work are necessarily optimal in the sense of achieving the smallest possible estimation error, but rather they are a trade-off between accuracy and susceptibility to non-optimal Toeplitz structure.

\section{Results}
In this section we will provide numerical results for some of the setups and methods discussed above. Note that the CRB for the complete system $\text{CRB}\left(\bm{\theta}\right)$ is approximately $M$ times lower than \eqref{eq:crb_nfo} and \eqref{eq:crb_fo} for the single oscillator case due to the fact that there are essentially $M$ observations of the same parameters, as partly discussed in \cite{Besson2003,Cheng2013}.

\subsection{Frequency Offset free Case - single Oscillator}
\begin{figure}[!t]
\centering
\columnplot
%
%
%
\definecolor{mycolor1}{rgb}{0.00000,0.44700,0.74100}%
\definecolor{mycolor2}{rgb}{0.85000,0.32500,0.09800}%
\definecolor{mycolor3}{rgb}{0.92900,0.69400,0.12500}%
\definecolor{mycolor4}{rgb}{0.49400,0.18400,0.55600}%

\definecolor{usualBlue}{rgb}{0.0784313753247261,0.168627455830574,0.549019634723663}
\begin{tikzpicture}

\begin{axis}[%
width=\figurewidth,
height=\figureheight,
scale only axis,
xmin=0,
xmax=30,
xlabel={SNR in \SI{}{\decibel}},
xmajorgrids,
ymode=log,
ymin=0.0001,
ymax=1,
yminorticks=true,
ylabel={MSE of $h_{mn,\phi}$},
ymajorgrids,
yminorgrids,
legend style={at={(1.00,1.00)},legend columns=1,anchor=north east,draw=black,fill=white,legend cell align=left,font=\scriptsize}
]

\addplot [color=black,dashed]
  table[row sep=crcr]{0	0.055555556\\
3	0.027843735\\
6	0.013954925\\
9	0.00699403\\
12	0.003505319\\
15	0.001756821\\
18	0.000880496\\
21	0.000441293\\
24	0.000221171\\
27	0.000110848\\
30	5.56e-05\\
};
\addlegendentry{CRB};

\addplot [color=black,dashed,mark=square]
  table[row sep=crcr]{0	0.501229598\\
3	0.250199044\\
6	0.125614857\\
9	0.063057684\\
12	0.031524063\\
15	0.015795291\\
18	0.007896955\\
21	0.003971092\\
24	0.00199113\\
27	0.000998758\\
30	0.000499969\\
};
\addlegendentry{Standard LS};	

\addplot [color=usualBlue,solid,mark=x]
  table[row sep=crcr]{0	0.49923185\\
3	0.2516398\\
6	0.125841842\\
9	0.063089606\\
12	0.031415894\\
15	0.015835641\\
18	0.007911076\\
21	0.003959751\\
24	0.001985596\\
27	0.000998008\\
30	0.00050243\\
};
\addlegendentry{Toep., $P=1$};

\addplot [color=usualBlue,solid,mark=o]
  table[row sep=crcr]{0	0.223603354\\
3	0.111150254\\
6	0.055825656\\
9	0.027970869\\
12	0.013852421\\
15	0.007020543\\
18	0.003538443\\
21	0.001766018\\
24	0.00088075\\
27	0.000442514\\
30	0.000220988\\
};
\addlegendentry{Toep., $P=3$};

\addplot [color=usualBlue,solid,mark=triangle]
  table[row sep=crcr]{0	0.118967064\\
3	0.05956286\\
6	0.030034599\\
9	0.015032005\\
12	0.007524501\\
15	0.003794943\\
18	0.001895124\\
21	0.000951171\\
24	0.000473654\\
27	0.000238589\\
30	0.000118242\\
};
\addlegendentry{Toep., $P=6$};

\addplot [color=usualBlue,solid]
  table[row sep=crcr]{0	0.073717864\\
3	0.036935577\\
6	0.018693051\\
9	0.009291927\\
12	0.004705678\\
15	0.002360855\\
18	0.001173266\\
21	0.000589152\\
24	0.000296273\\
27	0.000147244\\
30	7.38e-05\\
};
\addlegendentry{Toep., $P=9$};

\end{axis}
\end{tikzpicture}%
\caption{MSE for the estimation of the channel matrix based on a URA with $M_x=M_y=N_x=N_y=3$, i.e., $M=N=9$ , in a single oscillator setup without frequency offset.}
\label{fig:MSE_h_only}
\vspace{-1mm}
\end{figure}
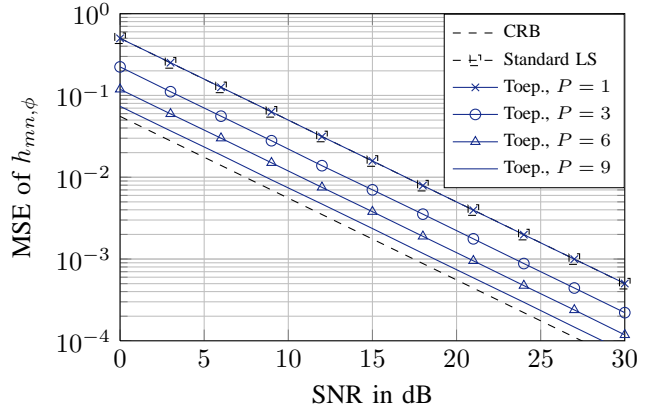

In Fig.~\ref{fig:MSE_h_only} we show the mean~squared~error~(MSE) results for an ideally designed URA with a single oscillator setup having a random initial phase offset $\Delta\phi$. To get the channel estimates we use the training matrix $\matr{X}=\matr{I}_N$, but truncated after $P$ rows. Then, the received vectors $\matr{y}_m$ contain the block Toeplitz structured rows/columns of the channel matrix. We average the equal entries within the sub-matrices and across equal sub-matrices to improve the accuracy of the estimates.

As can be seen, the performance exploiting the Toeplitz structure is at least as good as the least-squares~(LS) standard solution \cite{Biguesh2006}, which does not take the similarity between matrix rows/columns into account. By increasing the number of training vectors the MSE decreases due to averaging of noise as expected. The reason why the performance with $P=9$ training vectors is not equal to the CRB is that for the results shown here, only the structure in the blocks itself and across equal blocks is exploited. For this specific setup there are, however, also certain channel matrix entries that are equal over different sub-matrices $\matr{H}_{n_y}$.

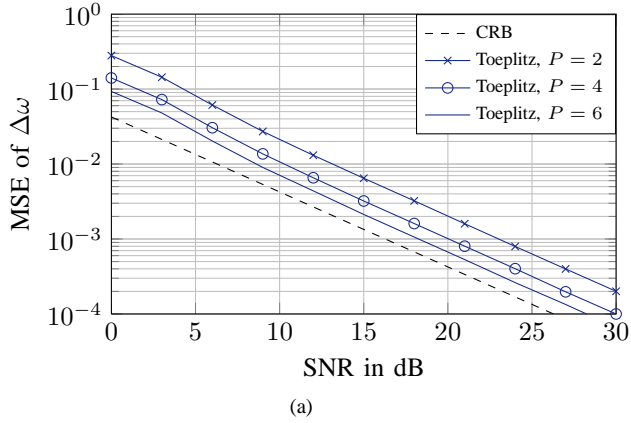
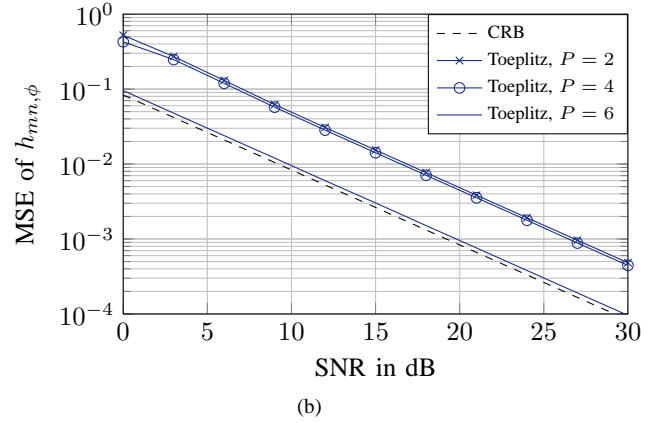
\begin{figure*}[!t]
\centering
\subfloat[]{\columnplottwo
%
%
%
\definecolor{mycolor1}{rgb}{0.00000,0.44700,0.74100}%
\definecolor{mycolor2}{rgb}{0.85000,0.32500,0.09800}%
\definecolor{mycolor3}{rgb}{0.92900,0.69400,0.12500}%
\definecolor{mycolor4}{rgb}{0.49400,0.18400,0.55600}%

\definecolor{usualBlue}{rgb}{0.0784313753247261,0.168627455830574,0.549019634723663}
\begin{tikzpicture}

\begin{axis}[%
width=\figurewidth,
height=\figureheight,
scale only axis,
xmin=0,
xmax=30,
xlabel={SNR in \SI{}{\decibel}},
xmajorgrids,
ymode=log,
ymin=0.0001,
ymax=1,
yminorticks=true,
ylabel={MSE of $\Delta\omega$},
ymajorgrids,
yminorgrids,
legend style={at={(1.00,1.00)},anchor=north east,draw=black,fill=white,legend cell align=left,font=\scriptsize}
]

\addplot [color=black,dashed]
  table[row sep=crcr]{0	0.042463144\\
3	0.021281986\\
6	0.010666259\\
9	0.005345793\\
12	0.002679243\\
15	0.001342803\\
18	0.000673\\
21	0.000337\\
24	0.000169\\
27	8.47e-05\\
30	4.25e-05\\
};
\addlegendentry{CRB};

\addplot [color=usualBlue,solid,mark=x]
  table[row sep=crcr]{0	0.2795595\\
3	0.14387909\\
6	0.061267881\\
9	0.027198319\\
12	0.013098488\\
15	0.006461505\\
18	0.003231758\\
21	0.001599478\\
24	0.000795847\\
27	0.000398838\\
30	0.000200761\\
};
\addlegendentry{Toeplitz, $P=2$};

\addplot [color=usualBlue,solid,mark=o]
  table[row sep=crcr]{0	0.140588181\\
3	0.07238949\\
6	0.03050931\\
9	0.0136598\\
12	0.006558401\\
15	0.003212213\\
18	0.001609077\\
21	0.00080258\\
24	0.00040174\\
27	0.000197475\\
30	0.0001\\
};
\addlegendentry{Toeplitz, $P=4$};

\addplot [color=usualBlue,solid]
  table[row sep=crcr]{0	0.092850049\\
3	0.047991302\\
6	0.020288481\\
9	0.009003163\\
12	0.004386167\\
15	0.002126708\\
18	0.001061674\\
21	0.000531856\\
24	0.000262115\\
27	0.000133974\\
30	6.61e-05\\
};
\addlegendentry{Toeplitz, $P=6$};

\end{axis}
\end{tikzpicture}%
\label{fig:MSE_w}}
\hfill
\subfloat[]{\columnplottwo
%
%
%
\definecolor{mycolor1}{rgb}{0.00000,0.44700,0.74100}%
\definecolor{mycolor2}{rgb}{0.85000,0.32500,0.09800}%
\definecolor{mycolor3}{rgb}{0.92900,0.69400,0.12500}%
\definecolor{mycolor4}{rgb}{0.49400,0.18400,0.55600}%

\definecolor{usualBlue}{rgb}{0.0784313753247261,0.168627455830574,0.549019634723663}
\begin{tikzpicture}

\begin{axis}[%
width=\figurewidth,
height=\figureheight,
scale only axis,
xmin=0,
xmax=30,
xlabel={SNR in \SI{}{\decibel}},
xmajorgrids,
ymode=log,
ymin=0.0001,
ymax=1,
yminorticks=true,
ylabel={MSE of $h_{mn,\phi}$},
ymajorgrids,
yminorgrids,
legend style={at={(1.00,1.00)},anchor=north east,draw=black,fill=white,legend cell align=left,font=\scriptsize}
]

\addplot [color=black,dashed]
  table[row sep=crcr]{0	0.083333333\\
3	0.041765603\\
6	0.020932387\\
9	0.010491045\\
12	0.005257978\\
15	0.002635231\\
18	0.00132\\
21	0.000662\\
24	0.000332\\
27	0.000166\\
30	8.33e-05\\
};
\addlegendentry{CRB};

\addplot [color=usualBlue,solid,mark=x]
  table[row sep=crcr]{0	0.521448868\\
3	0.271698567\\
6	0.130265343\\
9	0.062171934\\
12	0.030762185\\
15	0.015361157\\
18	0.007687626\\
21	0.003835043\\
24	0.001919396\\
27	0.000961925\\
30	0.000482196\\
};
\addlegendentry{Toeplitz, $P=2$};

\addplot [color=usualBlue,solid,mark=o]
  table[row sep=crcr]{0	0.427037224\\
3	0.247308579\\
6	0.119276329\\
9	0.057359921\\
12	0.028318279\\
15	0.014140171\\
18	0.007087361\\
21	0.003543231\\
24	0.001779737\\
27	0.000880983\\
30	0.000445\\
};
\addlegendentry{Toeplitz, $P=4$};

\addplot [color=usualBlue,solid]
  table[row sep=crcr]{0	0.095466693\\
3	0.047730853\\
6	0.023924783\\
9	0.012025899\\
12	0.005998048\\
15	0.003028409\\
18	0.001510148\\
21	0.000759613\\
24	0.000380479\\
27	0.000190658\\
30	9.58e-05\\
};
\addlegendentry{Toeplitz, $P=6$};

\end{axis}
\end{tikzpicture}%
\label{fig:MSE_h}}
\caption{MSE for the consecutive estimation of the frequency offset and channel for a ULA with $M=N=6$ in a single oscillator setup: \protect\subref{fig:MSE_w}~Frequency offset; \protect\subref{fig:MSE_h}~Channel coefficients.}
\label{fig:MSE_joint}
\end{figure*}

\subsection{Frequency Offset - single Oscillator}

Fig.~\ref{fig:MSE_joint} shows the MSE for a ULA with single oscillator setup with random initial phase offset and Gaussian distributed frequency offsets with variance of $\SI{0.3}{rad^2\per sample}$. ULAs are used here so that \eqref{eq:freq_est} can be applied. For block Toeplitz structures the same strategy can be used but it needs to be applied on each of the sub-matrices $\matr{H}_{n_y}$ separately.

We start with the estimation of the frequency offset by eliminating the impact of the channel gains from the receive vectors of neighboring antenna pairs as shown in \eqref{eq:freq_est}. Consecutively, the estimated frequency offset is removed from the received vectors and the Toeplitz structure is exploited to gain the channel estimates as in the case without frequency offset.

A comparison to other techniques from the literature is difficult as they need longer training sequences, usually $P>N$. Nevertheless, the CRBs should be a good indicator for the performance of the discussed method. The results show that with the minimum number of training vectors $P=2$ the parameters can be reasonably well estimated and that with $P=N$ results close to the CRB are achievable.

\begin{figure}[!t]
\centering
\columnplottwo
%
%
%
\definecolor{mycolor1}{rgb}{0.00000,0.44700,0.74100}%
\definecolor{mycolor2}{rgb}{0.85000,0.32500,0.09800}%
\definecolor{mycolor3}{rgb}{0.92900,0.69400,0.12500}%
\definecolor{mycolor4}{rgb}{0.49400,0.18400,0.55600}%

\definecolor{usualBlue}{rgb}{0.0784313753247261,0.168627455830574,0.549019634723663}
\begin{tikzpicture}

\begin{axis}[%
width=\figurewidth,
height=\figureheight,
scale only axis,
xmin=5,
xmax=30,
xlabel={Number of antennas $N$},
xmajorgrids,
ymode=log,
ymin=0.00001,
ymax=0.01,
yminorticks=true,
ylabel={MSE of $\Delta\omega$},
ymajorgrids,
yminorgrids,
legend style={at={(1.00,1.00)},anchor=north east,draw=black,fill=white,legend cell align=left,font=\scriptsize}
]

\addplot [color=black,dashed]
  table[row sep=crcr]{4	0.000768\\
8	0.000120124\\
12	3.79e-05\\
16	1.65e-05\\
20	8.59e-06\\
24	5e-06\\
28	3.28e-06\\
32	2.22e-06\\
};
\addlegendentry{CRB};

\addplot [color=usualBlue,solid,mark=x]
  table[row sep=crcr]{4	0.00332722\\
8	0.000719261\\
12	0.000306518\\
16	0.000165815\\
20	0.000105659\\
24	7.24e-05\\
28	5.33e-05\\
32	4.03e-05\\
};
\addlegendentry{Toeplitz, $P=N/2$};

\addplot [color=usualBlue,solid]
  table[row sep=crcr]{4	0.001676784\\
8	0.000359505\\
12	0.000153095\\
16	8.34e-05\\
20	5.28e-05\\
24	3.7e-05\\
28	2.66e-05\\
32	2.02e-05\\
};
\addlegendentry{Toeplitz, $P=N$};


\addplot [color=usualBlue,dashed]
  table[row sep=crcr]{4	0.003417106\\
8	0.001125624\\
12	0.000647054\\
16	0.000449238\\
20	0.000345771\\
24	0.000284003\\
28	0.000233124\\
32	0.000201097\\
};
\addlegendentry{Toep. Impaired, $P=N$};

\end{axis}
\end{tikzpicture}%
\caption{MSE for estimation of the frequency offset for ULAs with different numbers of antennas in a single oscillator setup, $M=N$ and $\text{SNR}=\SI{20}{\decibel}$.}
\label{fig:MSE_w_ants}
\vspace{-1mm}
\end{figure}
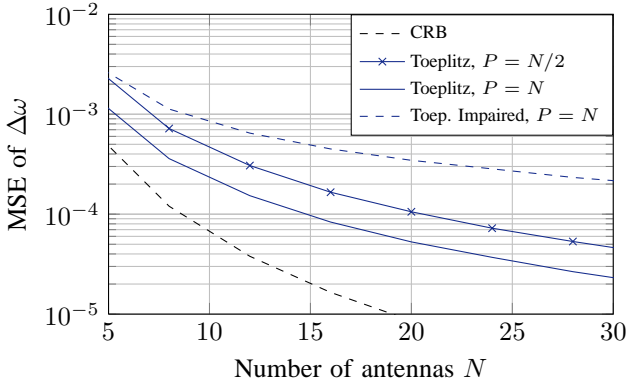

In Fig.~\ref{fig:MSE_w_ants} the results of the frequency offset estimation is shown versus the number of antennas. As expected, the estimation accuracy improves with $N$ as there are more observations of the same parameter. We have also added a case where the uniform antenna arrangement, i.e. Toeplitz structure, is impaired by small random positioning errors of the elements in all three possible array dimensions.

\section{Conclusion}
In this paper we have discussed how parameter estimation can be performed for pure LOS MIMO channels. It was shown that antenna arrangements with regular polyhedral structure generate channel matrices that are fully or partly of block Toeplitz structure. This structure is preserved if single oscillator setups are considered at transmitter and receiver. We showed that by exploiting this fact, very short training sequences are sufficient to gain accurate parameter values with simple estimators. This can be very useful to reduce training overhead for arrays with a very high number of densely packed antennas, where neighboring antennas will naturally experience highly dependent channels. Additionally, if longer training sequences are used the structure of the channel can be used in order to improve the accuracy of the estimates.

Relevant extensions of this work are the consideration of time varying the phase shifts \cite{Halsig2015a}, which can be partly dealt with by the methods presented here, and the consideration of setups where certain antenna groups have a shared oscillator, for which a Toeplitz structure may also exist. Furthermore, it is of interest how the geometrical dependencies that constitute the channel can be exploited in a more general framework.

\bibliographystyle{IEEEtran}
\bibliography{references}
%

\end{document}